\documentclass[]{epl}

\usepackage{amsmath}
\usepackage{amsfonts}
\usepackage{amssymb}
\usepackage{amscd}
\usepackage{graphicx}
\usepackage{psfrag}

\title{Jarzynski equality for the Jepsen gas}
\author{I. Bena\inst{1} \and C. Van den Broeck\inst{2} 
\and R. Kawai\inst{3}}
\institute{
  \inst{1} Department of Theoretical Physics, 
  University of Geneva, 1211 Geneva, Switzerland\\
  \inst{2} Hasselt University, B-3590 Diepenbeek, Belgium\\
  \inst{3} Department of Physics, University of Alabama at
Birmingham, AL 35294, USA }
\pacs{02.50.-r}{Probability theory, stochastic processes, and statistics}
\pacs{05.70.Ln}{Nonequilibrium and irreversible thermodynamics}
\pacs{05.20.-y}{Classical statistical mechanics}
\begin{document}

\maketitle

\begin{abstract}
We illustrate the Jarzynski equality on the exactly solvable model of a 
one-dimensional ideal gas in uniform expansion or compression. The analytical
results for the probability density $P(W)$ of the work $W$ performed 
by the gas are compared with the results of molecular dynamics simulations 
for a two-dimensional dilute gas of hard spheres. 
\end{abstract}

Exactly solvable models play an important role in statistical mechanics.
They complement and verify results that are derived from general and usually
abstract arguments,
while at the same time they offer insight and intuition.
They are particularly useful in far-from-equilibrium situations, 
for which few generic exact results are available.
In this letter, we focus on a  remarkable result in nonequilibrium statistical
mechanics,
namely the {\em  Jarzynski equality}~\cite{jarzynski,experiments}, which 
has given rise to a certain amount of confusion about its validity and
interpretation~\cite{cohen}.
The Jarzynski equality relates the statistics of the
amount of work $W$ performed by a system in a nonequilibrium transition between 
two equilibrium states, to the difference in the free energies of these states.
More precisely, one has:
\begin{equation}
\langle \exp(\beta W) \rangle \,=\,\exp(-\beta \Delta F)\,,
\label{jar}
\end{equation}
where $W$ is the work delivered by the system upon varying 
an external control parameter following a specified schedule between an
initial and final value, starting from an initial state of the system sampled
from a canonical
distribution at temperature $T$ ($\beta^{-1}=k_BT$). $W$ is
a random variable due to the sampling of the initial state. The first 
surprise is that the above specified average $\langle ...\rangle$ with 
respect to $W$ is independent of the schedule according to which  
the control parameter is changed between
the specified initial and  final values. In particular it is  
independent of whether this schedule keeps the system close
to equilibrium (quasi-static transformation) or whether it entails large
deviations from
equilibrium.  As a consequence the average is expressed in terms of the 
difference $\Delta F$ in free energy of the canonical equilibrium 
states at temperature $T$  at the final and initial values of the control parameter, respectively. The second surprise 
is then that  the equilibrium quantity $\Delta F$ can be obtained by
an ensemble average over nonequilibrium measurements. Finally and foremost, 
the appearance of  an equality in far from equilibrium dynamics 
is very surprising.  In fact, the equality
(\ref{jar}) leads, upon application of the Jensen inequality, to the familiar
inequality, 
$\langle W \rangle \leqslant W_{rev}=-\Delta F$,
corresponding to the formulation of the second principle of thermodynamics 
for a system in contact with a heat bath.

Our purpose here is to verify and complement the discussion of the Jarzynski
equation 
by deriving the analytic expression of the probability density $P(W)$ for 
the work $W$ in a  system with Newtonian dynamics.  
We will consider the so-called  {\em Jepsen gas},
for which a number of other exact equilibrium 
and nonequilibrium results have been derived~\cite{jepsen}. The Jepsen gas
consists of
$N$ identical point particles of mass $m$ moving on a line 
and  undergoing perfectly elastic collisions.
Actually, since the speeds are merely exchanged upon collision, 
and the identity of the particles is irrelevant, the same model can 
represent an ideal gas in which particles do not interact with each other. 
As initial state, $t=0$, we consider a system at thermal equilibrium 
at temperature $T$, i.e., the particles are uniformly distributed in the 
interval  $[-L,0]$, and their velocities are randomly
and independently chosen from the Maxwellian distribution 
$\phi(u)=\displaystyle\sqrt{{m\beta}/{2\pi}}\,\exp(-\beta mu^2/2)$. The 
right hand side of the interval is formed by a piston of infinite mass,
which is moved according to a specified schedule. 
Although more complicated situations can be considered, 
we will concentrate on the case of a piston moving at a constant velocity 
$V$ (positive or negative, i.e., corresponding to gas expansion or 
compression, respectively),
from the initial position  $X=0$ to the final position  $X=Vt$.
A similar model has been discussed in  \cite{lua}. 
Whereas the latter paper  focuses on the case where the work distribution 
is dominated by correlated recollisions of a single particle with the piston, 
we will study  the so-called {\em thermodynamic limit} of an infinite system, 
with $L\rightarrow \infty$, $N \rightarrow \infty$
and fixed density $n=N/L$. In this limit, recollisions of particles with 
the piston are no longer possible (at least not for the schedule considered
here).  
Furthermore, as we will show below, a comparison with the results for real 
but dilute gases becomes relevant.

To evaluate $W$, we first note that a particle of velocity $u_i>V$ 
colliding with the piston will recoil 
with the velocity $u'_i=2V-u_i$. Hence 
there is an energy transfer in the interval $[0,t]$ from this particle to the
piston given by
$\Delta W_i=2mV(u_i-V) \theta(x_i+u_i t-V t)$,
where the Heaviside function expresses that
the collision between the particle (with initial position $x_i<0$ 
and velocity $u_i$) and the piston has to take place before  time $t$.
We conclude that the probability to have a total energy transfer $W$ from the
gas to the piston
during the time interval $[0,t]$ is given by:
\begin{eqnarray}
&&P(W)=\langle\delta(W-\sum_{j=1}^N\, \Delta W_i)\rangle
= \left<  \int_{-\infty}^{\infty}\frac{dk}{2\pi}
\exp\left\{ik\left[W-\sum_{j=1}^N\,\Delta W_j\right]\right\} \right>_0
\nonumber\\
&&=\int_{-\infty}^{\infty}\frac{dk}{2\pi}\,\exp(ikW)\,
\left < \exp(-ik\Delta W_j) \right >_0^N \nonumber \\
&&=\int_{-\infty}^{\infty}\frac{dk}{2\pi}\,\exp
\bigg\{ikW \bigg.
\bigg. - nt\,\int_{V}^{\infty}du\, (u-V)\phi(u)\bigg.
\bigg.\big[1-\exp\left(-2ikmV(u-V)\right)\big]\bigg\} \,.
\label{expr1}
\end{eqnarray}
The average $\langle \ldots \rangle_0$ is taken over the distribution 
of the initial positions and
velocities of the particles. Note that we have not taken into account 
recollisions of the particles with the piston, hence only the final line, 
in which the thermodynamic limit has been taken, gives the exact result for
$P(W)$.

\begin{figure}[tb]
\psfrag{v = -1.0}{{$v=-1.0$}}
\psfrag{v = -0.5}{{$v=-0.5$}}
\psfrag{v = -0.1}{{$\hspace{-0.4cm}v=-0.1$}}
\psfrag{v = 1.0}{{$\hspace{-0.2cm}v=1.0$}}
\psfrag{v = 0.5}{{$v=0.5$}}
\psfrag{v = 0.1}{{$\hspace{0.2cm}v=0.1$}}
\psfrag{w}{{$w$}}
\psfrag{P(w)}{{$P(w)$}}
\begin{center}
\includegraphics[width=7cm]{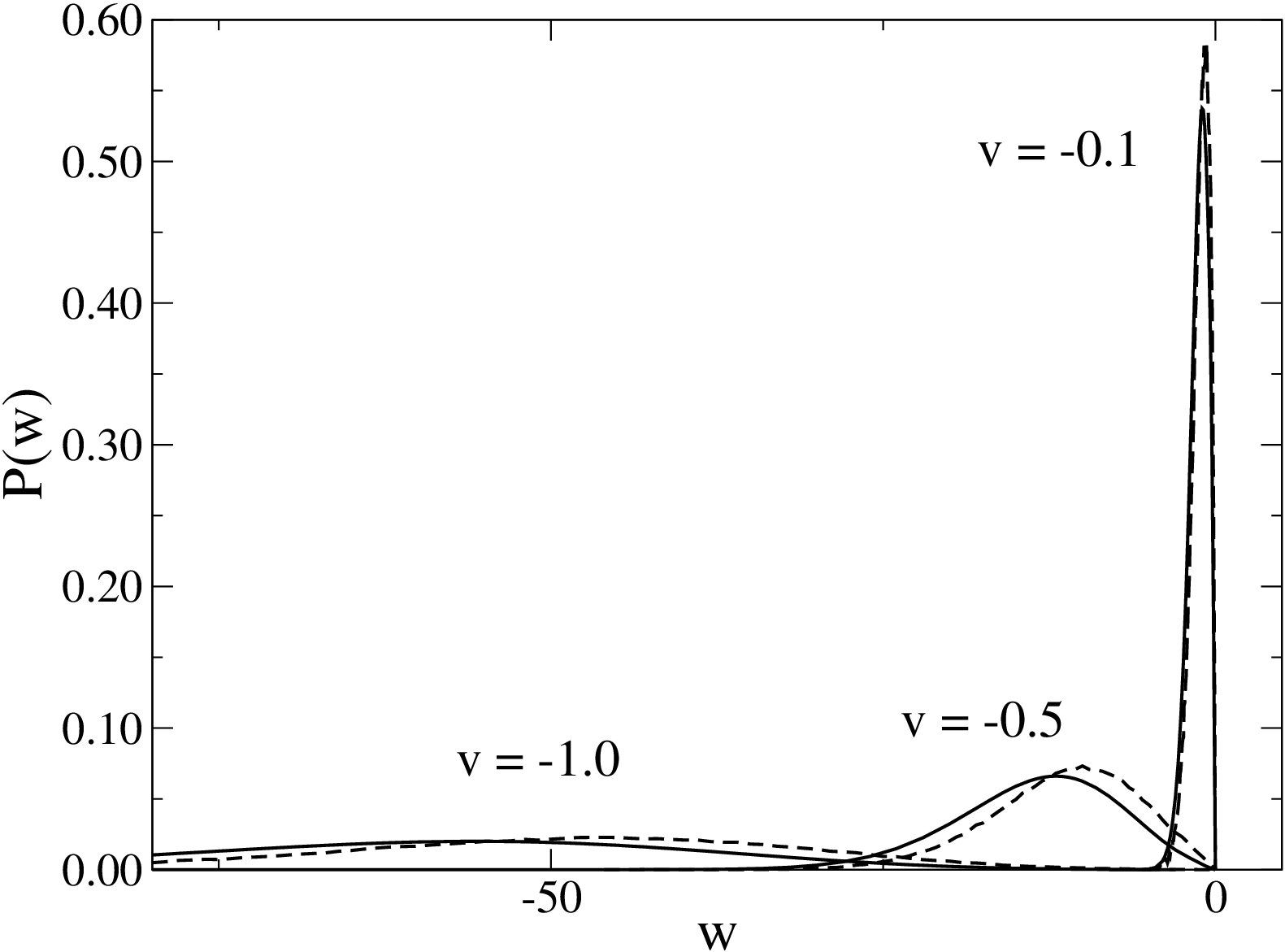}
\includegraphics[width=7cm]{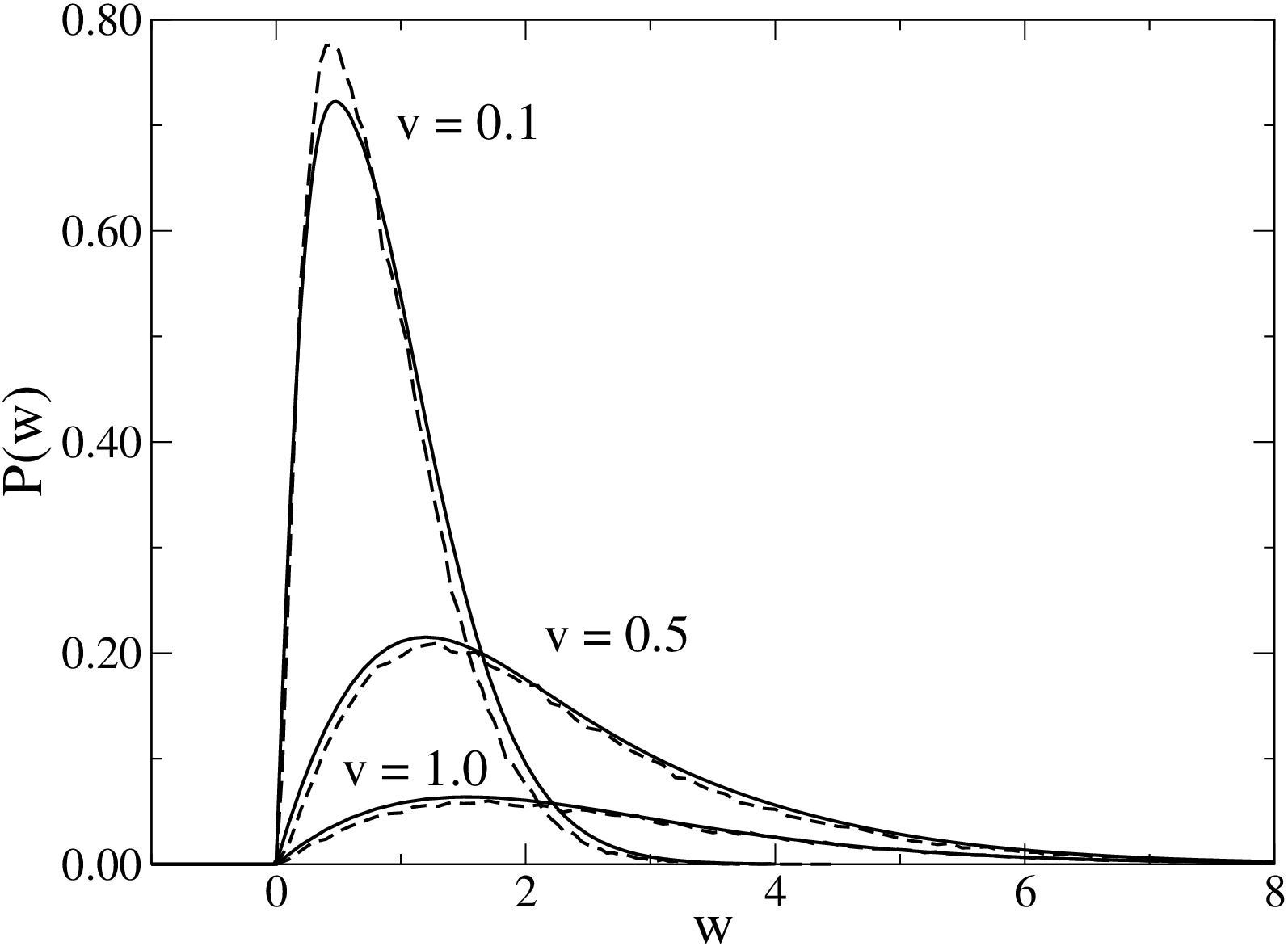}
\end{center}
\caption{Profiles of the nonsingular part of $P(w)$ for various velocities
$v$ of the piston at a fixed time $\tau=5$ 
(left: gas compression; right: gas expansion).
The area below the curves represents the probability
to have a nonzero value of $w$ and is equal to $[1-\exp(-\tau C_0)]$. 
The graphs illustrate the agreement between analytical calculations
(solid lines) and  molecular dynamics simulations results (dashed lines), see
the main text.}
\label{velocities}
\end{figure}

\begin{figure}[tb]
\psfrag{v = -1.0}{{$v=-1.0$}}
\psfrag{v = -0.5}{{$v=-0.5$}}
\psfrag{v = -0.1}{{$v=-0.1$}}
\psfrag{v = 1.0}{{$v=1.0$}}
\psfrag{v = 0.5}{{$v=0.5$}}
\psfrag{v = 0.1}{{$v=0.1S$}}
\psfrag{w}{{$w$}}
\psfrag{P(w)}{{$P(w)$}}
\begin{center}
\includegraphics[width=7.0cm]{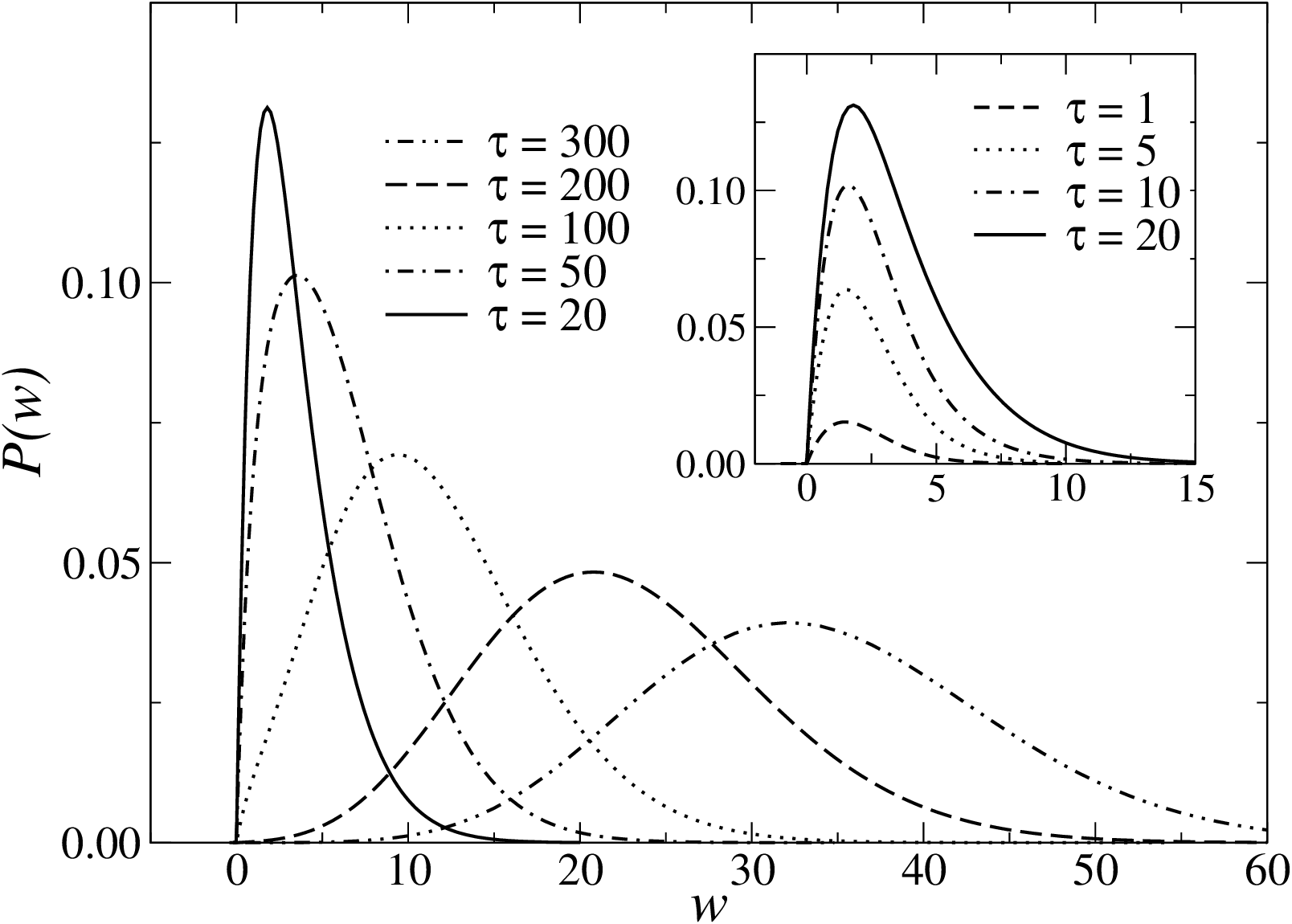}
\includegraphics[width=7.0cm]{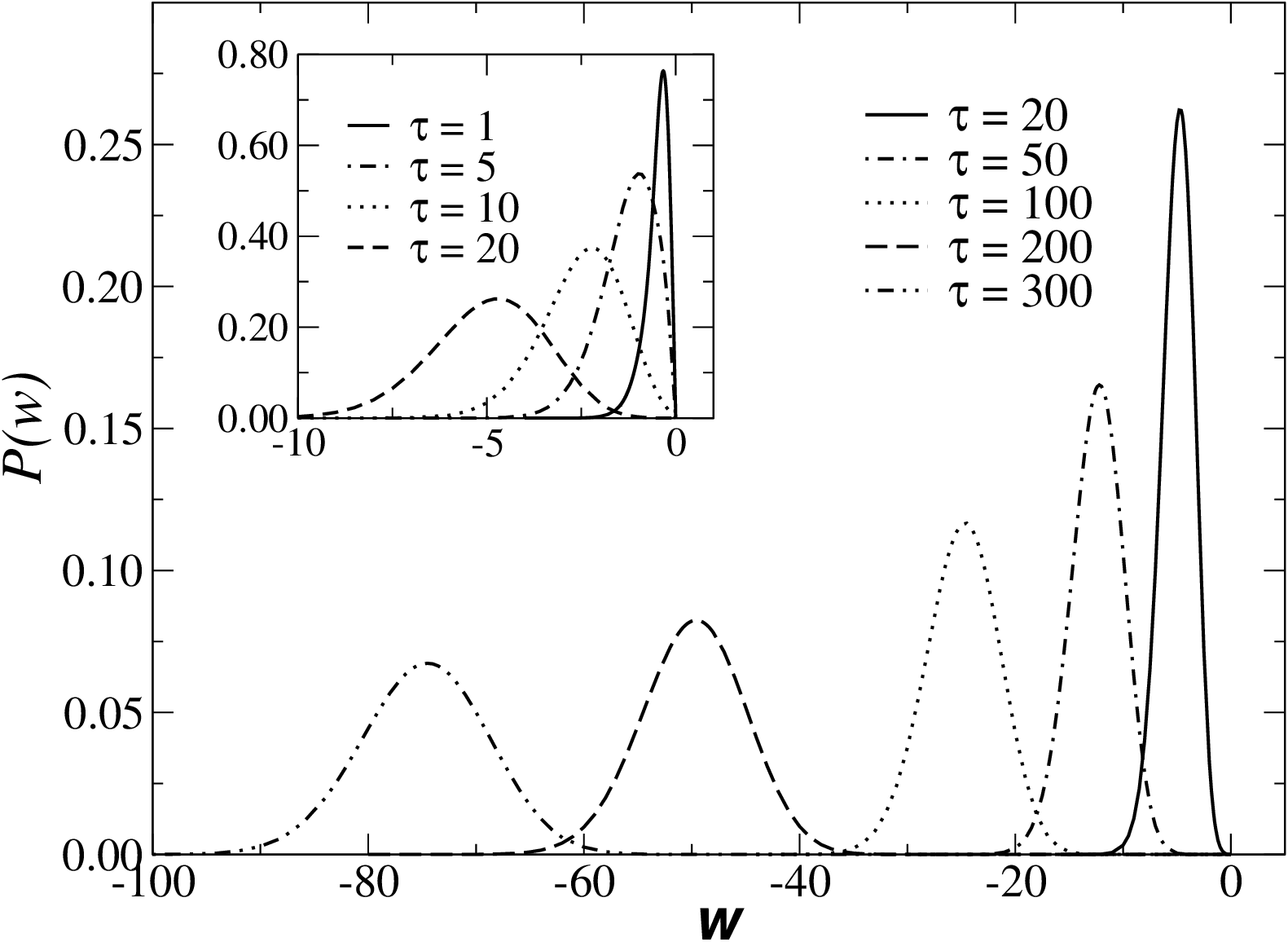}
\end{center}
\caption{Profiles of the nonsingular part of $P(w)$ at different times 
$\tau$ for fixed velocities of the piston, $v=1$ (left)
and $v=-0.1$ (right).
Note the evolution from a 
highly asymmetric profile at short times, 
Eq.~(\ref{short}), to an asymptotic Gaussian shape with 
mean and standard deviation given by Eqs.~(\ref{meanval})
and (\ref{variance}) respectively. }
\label{times}
\end{figure}

\begin{figure}[tb]
\psfrag{v}{{$v$}}
\psfrag{w}{{$w$}}
\psfrag{P(w)}{{$P(w)$}}
\begin{center}
\includegraphics[width=6.7cm]{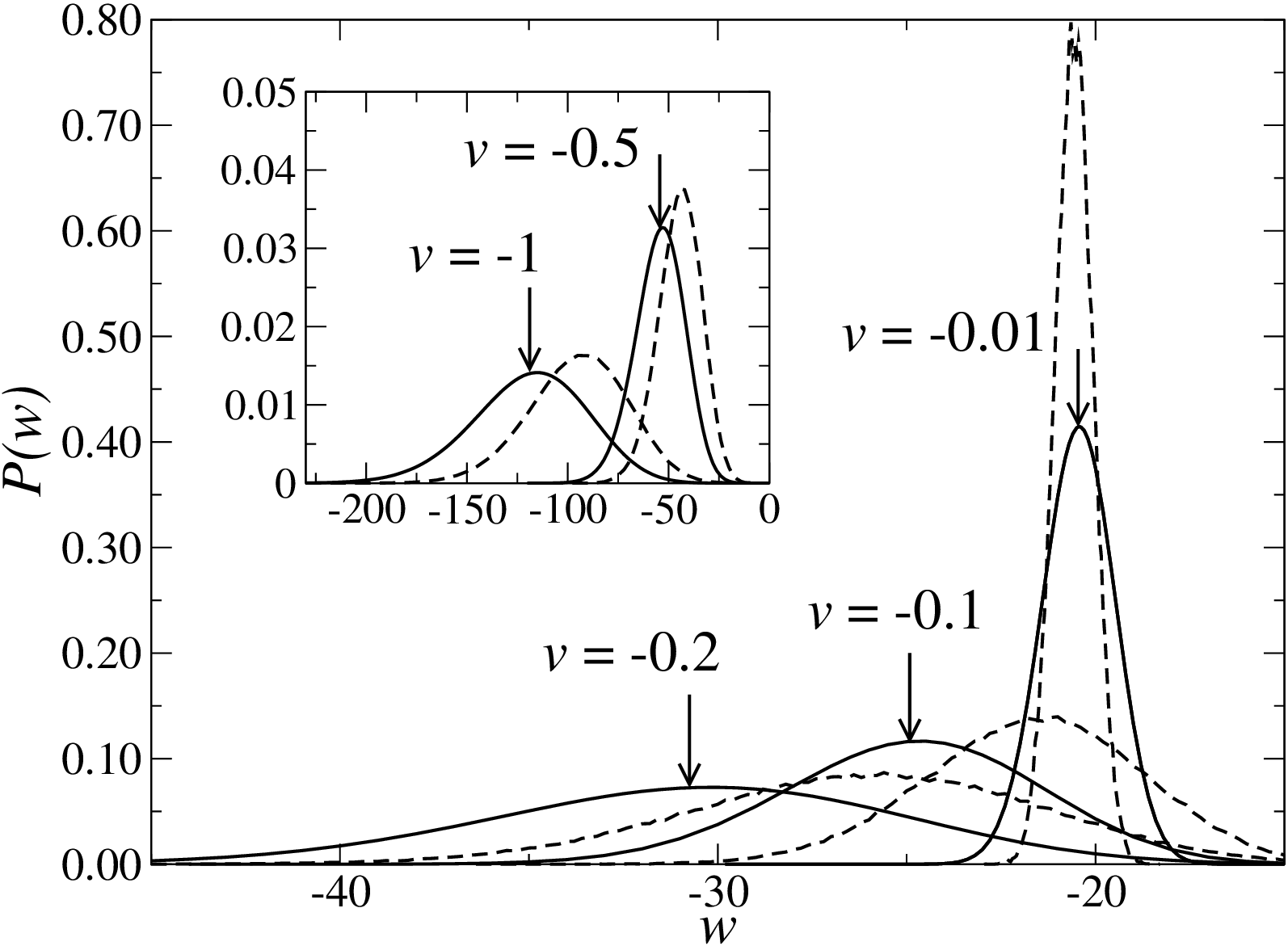}
\includegraphics[width=6.7cm]{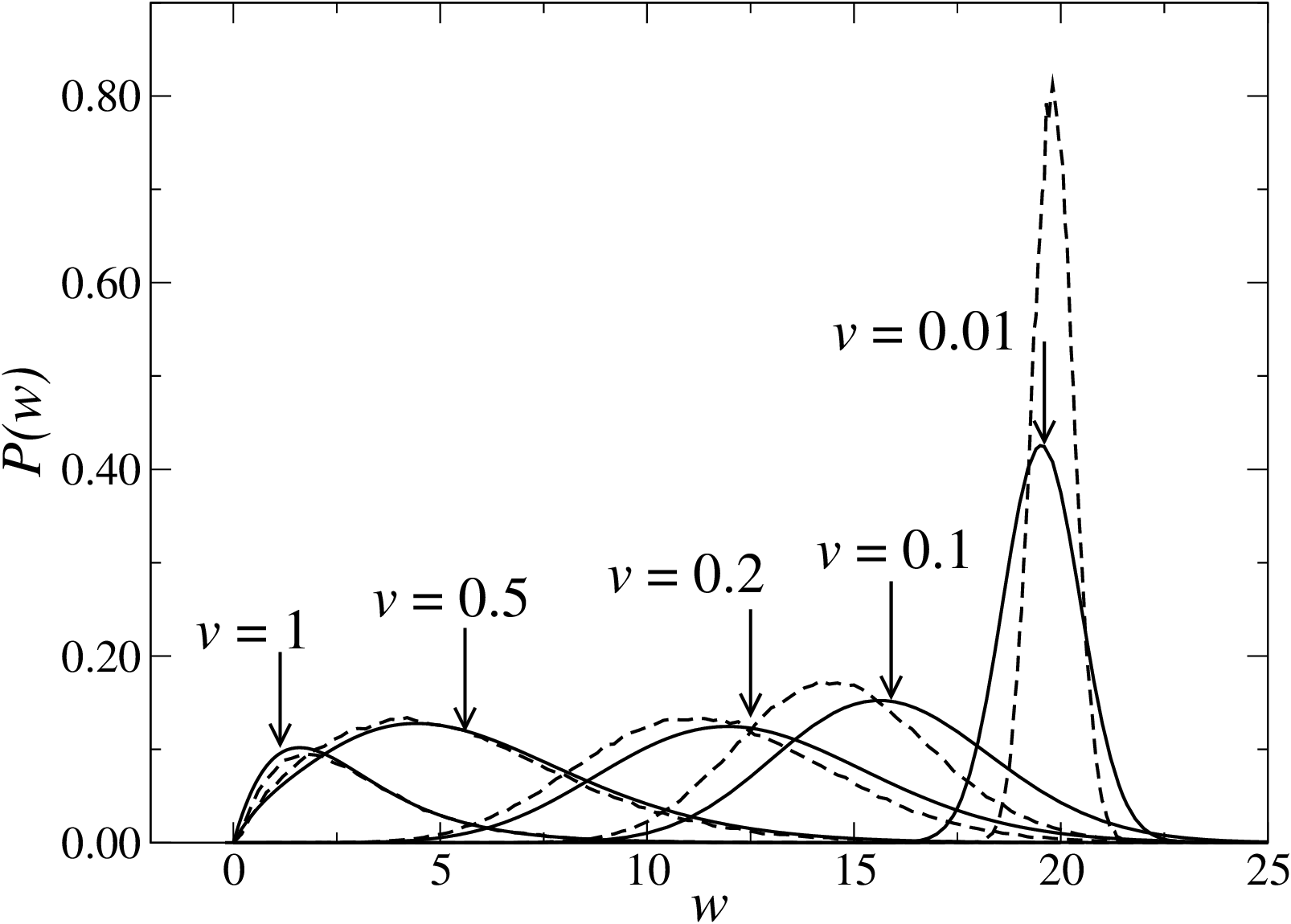}
\end{center}
\caption{Profiles of the nonsingular part of $P(w)$ for a fixed value
of $|w_{rev}|=2|v|\tau=20$ and different velocities $v$ of the piston (left:
compression, right: expansion). 
The inset shows the same plots for faster compression.
The arrows indicate the theoretical means, Eq.~(\ref{meanval}).
The theoretical results (solid lines) agree qualitatively  with  
the computer simulation (dashed lines). Quantitative disagreement 
for fast compression and slow expansion  are presumably due to recollisions and finite size effects.
}
\label{fixedvt}
%\vspace{0.5cm}
\end{figure}

The  schedule under consideration can be conveniently characterized by 
the following two  dimensionless variables:
\begin{eqnarray}
&&v=V ({\beta m/2})^{1/2},\quad\tau=nt/(2m\beta)^{1/2},
\end{eqnarray}
which are, essentially, the velocity of the piston 
measured in terms of the thermal
speed of the gas particles, 
and the average number of collisions during 
the considered time interval $[0,t]$ for a stationary piston. 
In terms of the scaled work $w=\beta W$,
the expression (\ref{expr1}) for the probability density becomes:
\begin{equation}
P(w)= \int_{-\infty}^{\infty}\frac{dq}{2\pi}\exp\left[iqw-\tau C(q)\right]\,,
\label{expr2}
\end{equation}
where the function $C(q)$ can be written as the sum of two parts, 
$C(q)=C_0+\tilde{C}(q)$.
The ``collisionless" part,
\begin{eqnarray} 
C_0 &=&{1}/{\sqrt{\pi}}\,\exp\left(-{v^2}\right)-
{v}\,\mbox{erfc}({v})\,,
\end{eqnarray}
(where $\mbox{erfc}(...)$ is the complementary error function)
corresponds to the absence of collisions between the gas 
particles and the piston, i.e., when no work is performed. 
This leads to a  singular contribution 
$\exp(-\tau C_0)\,\delta(w)$ in the expression 
(\ref{expr2}) of $P(w)$. This contribution is exponentially 
decaying with respect to the scaled time $\tau$, while
the damping exponent $C_0$ 
is a rapidly decaying function of $v$.
The second part,
\begin{eqnarray}
&&\tilde{C}(q)=
{v}(1+2iq)\,\mbox{erfc}\left[{v}
(1+2iq)\right]\exp\left[{v^2}(1+2iq)^2-{v^2}\right]-{1}/{\sqrt{\pi}}\,
\exp\left(-{v^2}\right)
\label{expr3}
\end{eqnarray}
is determined by the collisions between the gas particles and 
the piston, and leads to a mono-modal
nonsingular contribution to $P(w)$, as illustrated in 
Figs.~\ref{velocities}, \ref{times}, and \ref{fixedvt}. 

Based on this exact result, we first turn to the explicit 
verification of the Jarzynski equality.
One has:
\begin{eqnarray}\label{jarcheck}
&&\langle\exp(\beta W)\rangle=\langle
\exp(w)\rangle=\int_{-\infty}^{\infty}\frac{dq}{2\pi}\,\exp(-\tau C(q))\;
\int_{-\infty}^{\infty} dw \,\exp(iqw+w)\nonumber\\
&&=\exp(-\tau C(i)) = \exp(2\tau v)=\exp(ntV)\,.
\end{eqnarray}
On the other hand, the change in the
equilibrium free energy of the ideal gas due to the variation 
of its volume is a purely
entropic factor given by:
\begin{equation}
\exp(-\beta \Delta F)=\left(\frac{L+V t}{L}\right)^N 
\begin{CD}@>{\mbox{thermodynamic \;limit}}>>\end{CD} \exp(ntV)\,,
\end{equation}
so that  the Jarzynksi equality~(\ref{jar}) is indeed reproduced.

\begin{figure}
\begin{minipage}{6.9cm}
\psfrag{v}{{$v$}}
\psfrag{w}{{$w$}}
\psfrag{P(w)}{{$P(w)$}}
\psfrag{A}{$\hspace{-0.35cm}\langle w \rangle /\tau$}
\psfrag{B}{$\hspace{-1.1cm}\displaystyle\frac{\langle w \rangle-w_{rev}}{\tau}$}
\includegraphics[width=6.7cm]{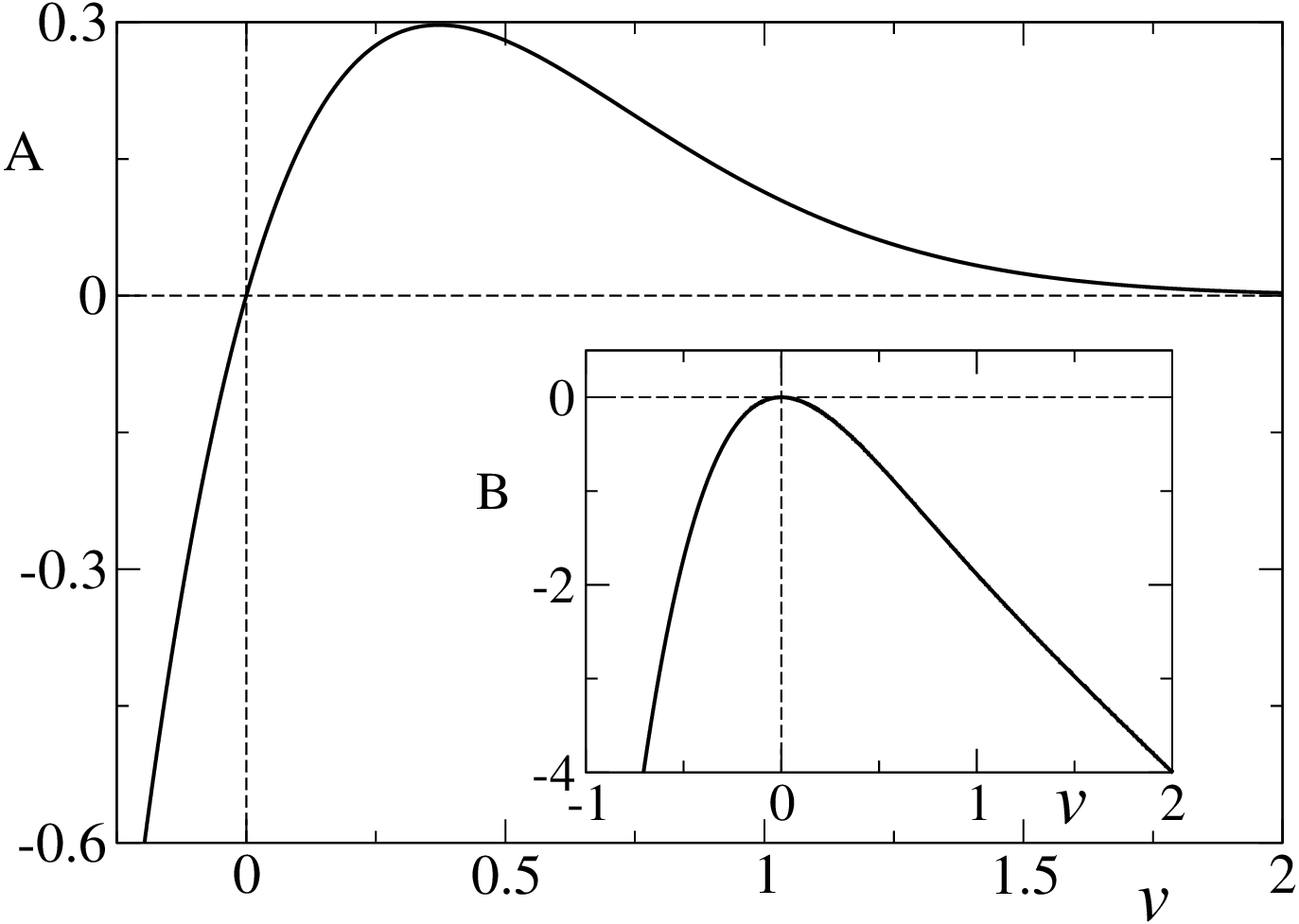}
\end{minipage}
\hspace{4mm}
\begin{minipage}{6.5cm}
\begin{tabular}{c|cccc}
\hline
{\it v} & mean & variance & skewness & kurtosis \\ \hline\hline
& 19.8 & 0.25 & 0.015 & 0.0016 \\ 
\raisebox{1.5ex}[0pt]{0.01} & (19.6) & (0.88) & (0.057) & (0.0036) \\ \hline
& 14.5 & 5.42 & 0.18 & 0.013 \\ 
\raisebox{1.5ex}[0pt]{0.1} & (15.9) & (6.88) & (0.20) & (0.044) \\ \hline
& 1.07 & 3.37 & 2.09 & 5.06 \\ 
\raisebox{1.5ex}[0pt]{1.0} & (1.13) & (3.50) & (2.03) & (4.81)\\ \hline\hline
& -20.60 & 0.26 & -0.014 & 0.0041 \\
\raisebox{1.5ex}[0pt]{-0.01} & (-20.46) & (0.93) & (-0.055) & (0.0035) \\ \hline
& -21.75 & 8.42 & -0.14 & 0.030 \\
\raisebox{1.5ex}[0pt]{-0.1} & (-24.92) & (11.71) & (-0.16) & (0.029) \\ \hline
& -97.47 & 585.8 & -0.25 & 0.044 \\
\raisebox{1.5ex}[0pt]{-1.0} & (-118.9) & (803.5) & (-0.27) & (0.077) \\ \hline

\end{tabular} 
\end{minipage}
\caption{The mean value of the energy transfer per unit time $\langle w
\rangle/\tau$, 
and, in inset, $(\langle w \rangle - w_{rev})/\tau$, both as a function of $v$.
 Note that $\langle w \rangle \leqslant w_{rev}$,
in agreement with the second law of thermodynamics. 
}
\label{mean}

\medskip
Table 1. $-$ Comparison of the mean values and higher moments obtained from the
molecular dynamics simulations 
with analytical results (values in parentheses), for different 
values of the velocity $v$ of the piston and a fixed value of 
$|w_{rev}|=2|v|\tau=20$. The
agreement is satisfactory, except for a few cases, 
cf. the discussion in the main text.
\end{figure}

We next  examine the characteristic properties of  $P(w)$. 
While an explicit evaluation of the Fourier transform~(\ref{expr2}) 
 appears to be difficult, one can easily obtain 
exact results for the moments using the  characteristic function 
of the probability density, namely
\begin{eqnarray}
G(q)=\langle \exp(-iqw)\rangle = \sum_{n=0}^{\infty}
\frac{(-iq)^n}{n!}\langle w^n\rangle=\exp(-\tau\,C(q))\,.
\label{moments}
\end{eqnarray}
In particular, one obtains the mean value  of the transferred energy
\begin{equation}
\langle w\rangle={2}v\tau\,\left[(1+2v^2)\mbox{erfc}({v})-
{2}/{\sqrt{\pi}}\,v\,\exp\left(-{v^2}\right)\right]\,,
\label{meanval}
\end{equation}
that increases linearly with time and has a maximum 
as a function of $v$ (see Fig.~\ref{mean}).
The centered moments of second through fourth order are 
\begin{eqnarray}
&&\sigma^2=\langle w^2\rangle-\langle w\rangle^2 = 8v^2\tau\,
\bigg[{2}/{\sqrt{\pi}}\,(1+v^2)\,\exp\left(-{v^2}\right)-v(3+2v^2)\mbox{erfc
}({v})\bigg]\,,
\label{variance} \\
&&\langle (w -\langle w \rangle)^3\rangle=16v^3\tau\,
\bigg[(3+12v^2+4v^4)\mbox{erfc}({v})\bigg.\bigg.-{2}/{\sqrt{{\pi}}}\,v\,(5+
2v^2)\,\exp\left(-{v^2}\right) \bigg]\,,
\label{mu3}\\
&&\langle (w -\langle w \rangle)^4\rangle=3 \sigma^4+64 v^4 \tau
\bigg[{2}/{\sqrt{\pi}}\,(4+9v^2+2v^4)\exp\left(-{v^2}\right) \bigg.
\nonumber\\ 
&&\bigg.\hspace{7cm}
- v\,(15 + 20 v^2 + 4v^4)
\text{erfc}({v})\bigg]\,.  
\label{mu4}
\end{eqnarray}

The explicit analytical expression
of  $P(w)$ can be derived in specific limits.

(i) In the {\em limit of large number of collisions} 
$\tau \gg 1$, the work is the sum of a large number of independent
contributions. One finds, in agreement with the central limit theorem, that
the distribution function  $P(w)$ converges to a Gaussian.  The 
mean and standard deviation 
are given by Eqs.~(\ref{meanval}) and (\ref{variance}).
The skewness 
$\gamma_3=\displaystyle{\langle (w -\langle w
\rangle)^3\rangle}/{\sigma^3}={\cal O}(\tau^{-1/2})$ 
and the 
kurtosis $\gamma_4=
\displaystyle{\langle (w -\langle w \rangle)^4\rangle}/{\sigma^4}-3=
{\cal O}(\tau^{-1})$ decay to zero.

(ii) For the {\em limit of  small number of collisions} $\tau \ll 1$, the moments
generating function reads
$G(q) \approx 1-\tau C(q)$, hence
\begin{eqnarray}
&&P(w) \approx (1 - \tau C_0) \delta (w)  + \frac{\tau}{8\sqrt{\pi} v^2} \;
|w|\; 
\exp\left[-\,\left(\frac{w}{4v}+v\right)^2\right]\;\theta(w v)\,,
\label{short}
\end{eqnarray}
corresponding to an asymmetric probability density profile (compare, for $v>0$,
with Eq.(18) from Ref.~\cite{lua} for the case of a single particle, $n=1/L$). 

(iii) The {\em quasi-static limit}. The reversible work $W_{rev}$ 
delivered by the
ideal gas during an expansion over a distance $\Delta x$ is $P \Delta x$,
with the pressure $P$ given by $n k_B T$. This result has to be compared
with the irreversible expansion at a finite speed $V$ over the same
distance, i.e., $\Delta x=Vt$. In terms of the previously introduced
dimensionless variables, this implies 
$w_{rev}=2 v \tau\,.$ This result is
indeed recovered in the limit 
$|v| \ll 1$, since $C(q) \approx {2} v iq$, and 
$P(w) \approx\delta(w-w_{rev})$.
By including a first order correction,
$C(q) \approx \displaystyle{8 v^2 q^2}/{\sqrt{\pi}}+{2}v iq \left(1 +
\displaystyle{4 v}/{\sqrt{\pi}}\right)$\,,
the probability density  $P(w)$ assumes a Gaussian profile, with mean and
variance:
\begin{equation}
\langle w \rangle \approx w_{rev} - \displaystyle{8 \tau
v^2}/{\sqrt{\pi}}\,,\quad
\langle w^2\rangle-\langle w\rangle^2 \approx \displaystyle{16 \tau
v^2}/{\sqrt{\pi}}\,.
\label{close}
\end{equation}

Note that  the Jarzynski equality Eq.~(\ref{jar}) implies that for
a Gaussian distribution the  fluctuation-dissipation ratio
\begin{equation}
{\cal R}=\frac{\langle w^2\rangle-\langle
w\rangle^2}{2(w_{rev}-\langle w\rangle)}
\label{R}
\end{equation}
is equal to $1$.
This is indeed the case in the
quasi-static limit, cf.~Eq. (\ref{close}), but {\em not} in the
long time limit, cf.  Eqs.~(\ref{meanval}) and (\ref{variance}).
The origin of this problem can be traced back to the contribution of large deviations to the Jarzynski average:
this average corresponds to the characteristic function evaluated 
at the complex unit, $G(q=i)$, 
cf. Eqs.~(\ref{jarcheck}) and (\ref{moments}). 
Even when higher order cumulants  converge to zero for 
the rescaled variable $(w-\langle w\rangle)/\sigma$ (central limit theorem), 
there is no guarantee that the contribution of the non-Gaussian 
tails can be neglected in the Jarzynski average.
Hence the application of the Jarzynski equality using 
a Gaussian ansatz is not reliable and can lead,
in a numerical or real experiment where a Gaussian distribution is observed,
to an erroneous value of the corresponding free energy difference.

%%%%%%%%%%%%%%%%%%%%%%%%%%%%%%%%%%%%%%%%%%%%%%%%%%%%%%%%%%%%%%%%%%%%%%%%%%%%%%%
%%%%%%%%%%%%%%%%%%%%%%%%%%%%%%%%%%%%%%%%%%%%%%%%%%%%%%%%%%%%%%%%%%%%%%%%%%%%%%%

We finally turn to  the practical and  experimental relevance of 
the Jarzynski equality.  
In particular one may wonder to which extent the above ideal gas 
results are representative of experimentally accessible measurements  
in a  dilute gas. 
As a first step in answering this question, we have performed 
extensive  molecular dynamics simulations of a dilute two-dimensional 
gas of hard disks. Note that such simulations allow to investigate 
parameter regions of high piston speed and short times which may be 
difficult to reach in experiment. 
The obtained simulation results 
represent 
an average over a half million runs, for a dilute hard disk gas with $N = 2000$ disks (diameter $d= 1$, mass $m = 1$).  
The initial positions and velocities of the disks are sampled from 
a microcanonical ensemble in a cylinder of length 
$L= 10^4$ and cross section  $S=10^2$ 
(i.e., initial gas density $\rho =0.002$),  
and initial  "temperature" $T = 1$.     
Extra caution was taken to reduce correlations between samples.
To compare the results with the one-dimensional model, 
the effective projected density $n=\rho S$ has to be used.

A systematic comparison between analytical and simulation results was performed
for a wide range of values of the speed $v$ and of time $\tau$, 
cf.  Figs.~\ref{velocities} and \ref{fixedvt} for an illustration.
Overall, the analytical  and the simulation results agree qualitatively. 
In particular, 
the progressive change in the general shape of the probability distribution 
from adiabatic to non-adiabatic regime is well reproduced by the numerics.  
The comparison of moments given in Table 1 confirms this agreement. 
There are, however, some notable discrepancies, 
namely for the mean and variance
for the compression case ($v<0$) 
and  the variance for the slow piston ($|v|=0.01$ in Table 1 and
Fig.~\ref{fixedvt}).  These deviations can be explained by the fact 
that the molecular dynamics simulation deviates in two basic assumptions
from the analytical model, namely the ideality of the gas and the 
thermodynamic limit.
Although a rather low density is used, recollisions of the gas particles 
with the piston are not negligible. In particular, they are clearly visibly 
in the simulations when the piston compresses the gas.
This non-ideality presumably causes the shift in the mean value 
and the variance for $v<0$.
For very slow piston motion, recollisions are not a major problem,
but the absence of 
the thermodynamic limit causes discrepancies between theory and simulations.
Since $\tau=1000$ for $|v|=0.01$, several hundreds to a thousand particles 
collide with the piston out of the limited number of N=2000.  
Clearly the absence of fast particles in the tail 
of the Maxwellian distribution will cause a narrowing of the observed
$P(w)$ distribution.  

In conclusion, the verification or exploitation of the Jarzynski equality 
to measure free energy differences appears to be quite intricate.  
Due to the role of extreme events, the gain in
computational time, when using a fast 
route between initial and final values of the control parameter, 
is outdone by the exponential increase in the required statistics 
\cite{delago}. 
The above explicit calculation and molecular dynamics simulations 
provide a dramatic example of this problem: even in the limit 
when a Gaussian distribution correctly describes the  whole 
probability mass (in the sense of the central limit theorem), 
extreme events may still be needed to correctly perform the 
Jarzynski average. In particular, the fluctuation-dissipation 
ratio~(\ref{R}) need not approach the value $1$ in a limit where 
$P(w)$ converges to a Gaussian distribution.

%=========================================================
\begin{acknowledgments} 
We thank the StochDyn program of the ESF for support.
I.B.  thanks F. Coppex  and acknowledges support of 
the Swiss National Science Foundation.
\end{acknowledgments}
%=========================================================

%=========================================================


\begin{thebibliography}{}

\bibitem{jarzynski} C. Jarzynski, Phys. Rev. Lett. {\bf 78}, 2690 (1997);
Phys. Rev. E {\bf 56}, 5018 (1997); 
Acta Phys. Pol. B {\bf 29}, 1609 (1998); 
J. Stat. Phys. {\bf 96}, 415 (1999);
in {\em Dynamics of Dissipation}, 
Eds. P. Garbaczewski
and R. Olkiewicz, Springer, Berlin (2002);
G. E. Crooks, J. Stat. Phys. {\bf 90}, 1481 (1998); 
Phys. Rev. E {\bf 60}, 2721 (1999);
Phys. Rev. E {\bf 61}, 2361 (2000); 
J. Kurchan, J. Phys. A {\bf 31}, 3719 (1998); 
T. Hatano, Phys. Rev. E {\bf 60}, R5017 (1999); 
O. Mazonka and C. Jarzynski, arXiv: cond-mat/9912121.
F. Ritort, Poincar\'e Seminar {\bf 2}, 195 (2003); 
S. Mukamel,  Phys. Rev. Lett. {\bf 90}, 170604 (2003); 
D.J. Evans, Mol. Phys. {\bf 101}, 1551 (2003); 
W. De Roeck and C. Maes, Phys. Rev. E {\bf 69}, 026115 (2004); 
U. Seifert, J. Phys. A {\bf 37}, L517 (2004); 
C. Jarzynski and D. K. Wojcik, Phys. Rev. Lett. {\bf 92}, 230602 (2004);
O. Narayan and A. Dhar, J. Phys. A {\bf 37}, 63 (2004);
A. Dhar, Phys. Rev. E {\bf 71}, 36126 (2005); 
A. Imparato and C. Peliti, Europhys. Lett. {\bf 69}, 643 (2005);
arXiv: cond-mat/0501576; V. Chernyak {\em et al}, 
Phys. Rev. E {\bf 71}, 025102(R) (2005).

\bibitem{experiments} J. Liphart {\em et al}, 
Science {\bf 292}, 733 (2001);  
J. Liphart {\em et al},
Science {\bf 296}, 1832 (2002);
G. Hummer and A. Szabo, Proc. Natl. Acad. Sci. USA {\bf 98}, 3658 (2001); 
F. Ritort {\em et al}, Proc. Natl. Acad. Sci. USA {\bf 99}, 13544 (2002);
J. Gore  {\em et al}, Biophys. J. {\bf 84}, 474 A, Suppl. S (2003);
Proc. Natl. Acad. Sci. USA {\bf 100}, 12564 (2003); 
F. Douarche {\em et al}, 
Europhys. Lett. {\bf 70}, 593 (2005); arXiv: cond-mat/0504465. 


\bibitem{cohen} E. G. D. Cohen and D. Mauzerall, 
J. Stat. Mech. P07006 (2004);
C. Jarzynski, reply, J. Stat. Mech. P09005 (2004).

\bibitem{jepsen} D. W. Jepsen, J. Math. Phys. {\bf 23}, 405 (1965);
J. Piasecki, J. Stat. Phys. {\bf 104}, 1145 (2001); 
V. Balakrishnan {\em et al},
Phys. Rev. E {\bf 65}, 031102 (2002).


\bibitem{lua} R. C. Lua
and A. Y. Grosberg, J. Phys. Chem. B {\bf 109}, 6805 (2005).


\bibitem{delago}
D. A. Hendrix and C. Jarzynski, J. Chem. Phys. {\bf 114}, 5964 (2001);
K. P. N. Murthy, arXiv: cond-mat/0104167;
C. Jarzynski, Phys. Rev. E {\bf 65}, 046122 (2002);
S. X. Sun, J. Chem. Phys. {\bf 118}, 5769 (2003);
E. Atilgan and S. X. Sun, J. Chem. Phys. {\bf 121}, 10392 (2004);
H. Oberhofen {\em et al}, J. Phys. Chem. B {\bf 109}, 6902 (2005).
 


\end{thebibliography}
\end{document}